\documentclass{aa}
\usepackage{graphicx}
\usepackage{natbib}
\usepackage{amsmath}
\usepackage{txfonts}

\begin{document}

\title{The metallicity range of variables in M3}

\author{Johanna Jurcsik }

   \institute{Konkoly Observatory of the Hungarian Academy of Sciences,
              P. O. Box  67, H-1525 Budapest, Hungary\\
              \email{jurcsik@konkoly.hu}
             }
   \date{Received; accepted}

\abstract{
The recently published spectroscopic metallicities of RR Lyrae
stars in M3 \citep{s01} though show a relatively wide range of the
[Fe/H] values, the conclusion that no metallicity spread is real has
been drawn, as
no dependence on either minimum temperature or period 
was detected. Comparing these spectroscopic metallicities with
[Fe/H] 
calculated from the Fourier parameters of the light curves of the variables
a correlation between the [Fe/H] values appears. 
As a consequence of the independence of the spectroscopic and
photometric metallicities, this correlation points to the reality 
of a metallicity spread.
The absolute magnitudes of these stars follow a similar trend 
along both the spectroscopic and photometric metallicities as the 
general $\mathrm{M_V - [Fe/H]}$ relation predicts,
which strengthens that the detected metallicity range is real. 
}

 \maketitle

\section{Introduction}

Nearly every globular cluster contains stars of homogeneous metallicity.
The most prominent exception is $\omega$ Cen with its considerably large,
$\sim$1~dex spread in the [Fe/H] content of giants and subgiants.
Slight [Fe/H] inhomogeneity in M92 \citep{king98,lang98} has also
been detected. Concerning the other elemental abundances, detailed chemical 
composition analysis of globular cluster stars have revealed that
complicate 
patterns of abundance inhomogeneity exist in many clusters. 

M3 is one of the reference clusters of the globular cluster metallicity scale.
Spectroscopically, [Fe/H] has been already derived from high
dispersion studies of about 10 giant stars \citep{bell,kraf,cav,kraft}
which do not show any evidence of a [Fe/H] spread. 
The latest result gives an average [Fe/H] value of $-1.5$~dex  
on a metallicity scale based on measurements of FeII lines in order to
avoid
 the non-negligible bias of
overionization effects in FeI lines \citep{kraft}. The first spectroscopic 
data on direct [Fe/H] measurements of the cluster RR Lyrae variables 
have been recently published by \citet[hereafter SPS]{s01}. 
 Using moderate resolution spectra taken with the Hydra 
multifiber spectrograph  they conclude that the RR Lyrae variables are
also uniform in composition as no dependence on either minimum temperature
or period can be found based on data of 29 RR Lyrae
variables. Although the scatter in  SPS's
[Fe/H] values 
(${\rm[Fe/H]_{mean}}=-1.43, s.dev.=\pm 0.12$ and
${\rm[Fe/H]_{mean}}=-1.21, s.dev.=\pm 0.22$ from FeI and FeII
lines, respectively) can be explained by uncertainties of
model parameters and equivalent widths, it does not exclude the possibility
of some real star to star [Fe/H] abundance differences either.
 
In \citet{jk96} it has been shown that the [Fe/H] of RRab
stars calculated from the period (P) and the $\phi_{31}$
Fourier phase-difference of the light curve
is as accurate as most of the spectroscopic results. 
With this formula the spectroscopic [Fe/H] of the 81 calibrating RRab stars
could be reproduced with 0.14~dex standard deviation.

In a preliminary investigation it has been  found that the 
metallicities of the RRab stars in M3 calculated according to this photometric 
method show larger spread than expected if the [Fe/H] of the variables 
is homogeneous \citep{bj00}. 

The aim of the present Letter is to check the reality of any possible [Fe/H] 
inhomogeneity of the M3 variables comparing their spectroscopic and 
photometric metallicity values.

\section{Spectroscopic and photometric [Fe/H]}

The two metallicity determination methods -- namely, a) spectroscopic;
b) utilizing light curve parameters (photometric) --, are completely
independent,
therefore any correlation between these  metallicities
would indicate that the observed metallicity spread is real.

\subsection{Spectroscopic metallicities}

During the last few years strong arguments were raised against metallicity
determinations using FeI
lines  \citep{la96,ti99,kraft} because of the non-negligible 
departures from LTE in metal-poor stellar atmospheres resulting
overionization of iron. Based on these results for the comparison purposes
we use exclusively the
FeII metallicities of SPS, which have unexpectedly large, $>0.6$~dex range.

The signal-to-noise ratio (S/N) of the spectroscopic observations
varies between 28 and 76 whereas the number of spectra (NS) of individual
stars
are between 10 and 41. For V9 both the S/N  and NS have one of the
lowest
values,
30 and 10, respectively. Therefore we have decided to omit this star from the
spectroscopic sample. The $\mathrm{[Fe/H]_{FeII}=-0.87~dex}$ seems to be
unreliably large
for this star based on the comparison of its light curve with a large sample
of RRab light curves compiled in \citet{j98}. 
A field RR Lyrae star (UW Gru) has been found to show light changes very 
similar to M3/V9.

\begin{figure}[hh!!!!!!!!]
\resizebox{\hsize}{!}{\includegraphics{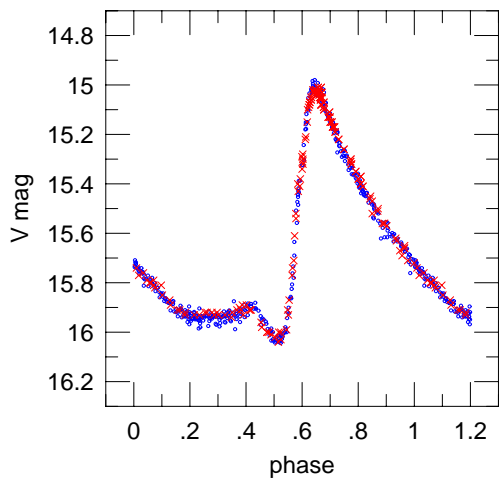}}
\vskip -2mm
\caption{Phased V light curves of M3/V9 (circles) and UW Gru
(crosses).
The magnitudes of UW Gru are shifted to match the
magnitudes of
M3/V9. The identity of the light curves' shapes indicates
close similarity
of the objects that have similar periods and colours.
Consequently, no significant metallicity
difference between these stars can be real.
\label{fig1}}
\end{figure}

Fig.~\ref{fig1} shows the phased light curves of M3/V9 and UW Gru using
the observations of \citet{cc01} and \citet{bj} for M3/V9, and \citet{bern} for UW ~Gru. 
The light curves of these two stars
can be hardly separated, within photometric uncertainties they can be
regarded as identical. We suppose that such equality of the light
curves with similar pulsation periods can only happen if there is
no significant difference in any of the basic physical properties of the two
stars. The periods of these variables are 
very similar and  their mean 
colours are also the same within the range limited by the photometric
uncertainties 
as given in Table~\ref{comp}.
The metallicity of UW Gru was given as $-1.68$~dex in \citep{layden}
which was transformed to  $-1.41$~dex metallicity value on a common [Fe/H] 
scale used in
\citet{jk96}. As a consequence of 
 the absolute equality of the light curves
of M3/V9 and UW Gru the spectroscopic $\mathrm{[Fe/H]=-0.87~dex}$ value
cannot
be 
correct for M3/V9. It  validates our decision to omit V9 from the
spectroscopic sample.

\begin{table}[!ht]
\caption[]{Parameters of M3/V9 and UW Gru  }\label{comp}
\begin{tabular}{ccccc}
\hline
\noalign{\smallskip}
                       & UW Gru & ref. & M3/V9& ref.\\
\noalign{\smallskip}
\hline
 \noalign{\smallskip}
 period [d]                          &  0.5482104       &     & 0.5415552   &   \\
$\mathrm{\langle B-V\rangle}$ [mag]  &   0.36          & 1   &   0.33, 0.39     & 2,3 \\
E(B$-$V) [mag]                       &  $0.10$      & 1   &   0.01& 4 \\ 
$\mathrm{[Fe/H]_{phot}}$ [dex]            & $-1.51 $         &     & $ -1.53$    &   \\
$\mathrm{[Fe/H]_{spect}}$ [dex]           & $-1.68$, $-1.41$ & 5,6 &$ -0.87$     & 7 \\  
\noalign{\smallskip}
\hline
\noalign{\smallskip}
\end{tabular}
\begin{list}{}{}
\item{
1) \citet{bern}; 2) \citet{bj}; 3) \citet{cc01}; 4) 
\citet{h96}; 5) \citet{layden}; 6) \citet{jk96}; 7) \citet{s01} }
\end{list}
\end{table}

\subsection{Photometric metallicities}

Reliable photometric [Fe/H] can be  determined for 19 RRab stars among 
the spectroscopic sample of SPS (29 stars). The following stars 
are omitted:
\begin{itemize}
\item{stars showing strong Blazhko modulation that makes 
[Fe/H]$_{\rm phot}$ incorrect (V20, V39, V52, V61, V63, V66, V78).}
\item{no or poor quality published light curves are available (V113, V115).}
\item{V202, which cannot be regarded as a normal RRab star based on its 
long period (0\fd77), light curve shape, and low amplitude
($A_V=0\fm18$).}
\end{itemize}
In order to increase the  small sample size
those Blazhko stars are also used in the comparison which
exhibit only slight light curve modulations (V10, V59 and V62).
The small amplitude modulations are supposed not to
affect significantly the calculated
photometric metallicities.

For most of the stars photometries are available from different
sources.
In these cases the data have been merged with 
necessary zero-point offsets ($0.01-0.05$~mag) and a single
 [Fe/H]$_{\rm phot}$
is calculated from the combined data set.
In each case the magnitude correction has been determined 
relative to the \citet{cc01} data. 
All the data have been cleaned from outlying points.
The large number of the photometric measurements (N$\approx120-700$)
makes the separation of discrepant data unambiguous in
most of the cases.
The formal errors of the [Fe/H]$_{\rm phot}$ values are between 0.03 and 0.07~dex.

\subsection{$\mathrm{[Fe/H]_{spect} - [Fe/H]_{ phot}}$}

The spectroscopic and photometric metallicities, the intensity averaged 
mean magnitudes according to \citet{cc01}, and
the used photometries
of the variables are listed in  Table~\ref{tbl-1}.

\begin{table}
\centering
\caption[]{Spectroscopic and photometric metallicities and
observed mean magnitudes of the SPS  RRab sample.} \label{tbl-1}
\begin{tabular}{rc@{\hspace{3.pt}}cc@{\hspace{5.pt}}c@{\hspace{5.pt}}l}
\hline
\noalign{\smallskip}
 Var.      & [Fe/H]$_{\rm spect}^{\mathrm a}$ &[Fe/H]$_{\rm phot}$&
$\langle V\rangle_{\rm int}$& Rem.$^{\mathrm b}$ & 
Ref. phot.$^{\mathrm c}$\\
&\multicolumn{2}{c}{dex}&mag&&\\
\noalign{\smallskip}
\hline
\noalign{\smallskip}
 9    &$-$0.87 &$-$1.53   &15.630&  *   &   2, 3 \\
10    &$-$1.55 &$-$1.48   &15.615& (Bl)    &   2, 3, 4, 5\\
15    &$-$1.24 &$-$1.47   &15.597&      &   2, 3,    5\\
18    &$-$1.06 &$-$1.30   &15.673&      &   2, 3,    5\\
20    &$-$1.51 &$-$       &15.611&  Bl  &   2, 3\\
36    &$-$1.43 &$-$1.44   &15.618&      &   2, 3\\
39    &$-$1.09 &$-$       &15.674&  Bl  &   2, 3\\
46    &$-$1.35 &$-$1.11   &15.668&      &1, 2, 3, 4\\
51    &$-$1.34 &$-$1.33   &15.621&      &   2, 3,    5\\
52    &$-$1.22 &$-$       &15.705&  Bl  &   2, 3,    5\\
59    &$-$1.06 &$-$1.23   &15.646& (Bl) &   2, 3,    5\\
60    &$-$1.28 &$-$1.28   &15.520&      &   2, 3\\
61    &$-$1.33 &$-$       &15.641&  Bl  &   2, 3,    5\\
62    &$-$1.28 &$-$1.14   &15.620& (Bl) &   2, 3,    5\\
63    &$-$0.83 &$-$       &15.661&  Bl  &   2, 3,    5\\
66    &$-$1.05 &$-$       &15.607&  Bl  &1, 2, 3, 4, 5\\
77    &$-$0.97 &$-$1.12   &15.725&  **    &1, 2, 3, 4\\
78    &$-$0.93 &$-$       &15.554&  Bl  &1, 2, 3, 4\\
81    &$-$1.41 &$-$1.43   &15.656&      &   2, 3\\
82    &$-$1.59 &$-$1.57   &15.601&      &      3\\
84    &$-$1.03 &$-$1.19   &15.630&      &1, 2, 3, 4, 5\\
90    &$-$1.05 &$-$1.36   &15.639&      &   2, 3,    5\\
93    &$-$1.60 &$-$1.37   &15.640&      &   2, 3\\
94    &$-$1.13 &$-$1.37   &15.677&      &   2, 3\\
113   &$-$1.48 &$-$       &$-$   &  **  &\\
114   &$-$0.92 &$-$1.27   &15.686&  $-$ &      3\\
115   &$-$1.30 &$-$       &$-$   &  **  &      3\\
116   &$-$1.42 &$-$1.26   &15.685&  $-$ &   2, 3\\
202   &$-$1.19 &$-$       &15.524&  *** &   2, 3\\
\noalign{\smallskip}
\hline
\end{tabular}

\begin{list}{}{}
\item[$^{\mathrm{a}}$]
[Fe/H]$_{\rm spect}$ = [Fe/H]$_{\rm FeII}$ from SPS.

\item[$^{\mathrm{b}}$]
Bl: no [Fe/H]$_{\rm phot}$ is calculated for Blazhko stars with
large light curve modulation;

(Bl): Blazhko stars with slight light curve changes are included but
photometric [Fe/H] is less certain;

* : spectroscopic [Fe/H] seems to be wrong, see the text for details;

** : photometric data have larger scatter;

*** : classification is uncertain;

$-$ : no or poor quality light curve is available.

\item[$^{\mathrm{c}}$] 
1) \citet{bj00}; 2) \citet{bj}; 3) \citet{cc01}; 4)
\citet{c98}; 5) \citet{k98}.
\end{list}
\end{table}

\begin{figure}[hhh]
\resizebox{\hsize}{!}{\includegraphics{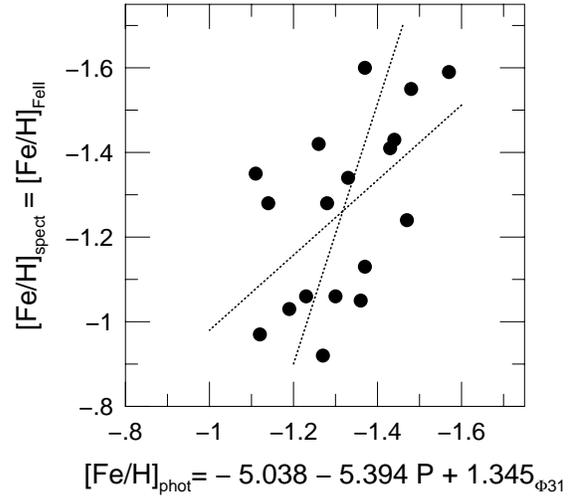}}
\caption{Spectroscopic vs. photometric [Fe/H] of 18 RRab stars in M3.
Dotted lines are the direct and inverse linear regression fits to the data. The
correlation
between the
spectroscopic and photometric metallicities seems to be significant.
As the two different measures of the metallicities are basically
independent, their correlation indicates that a metallicity range of
the M3 variables is indeed detected.
\label{fig2}}
\end{figure}

\begin{figure}[hhh]
\resizebox{\hsize}{!}{\includegraphics{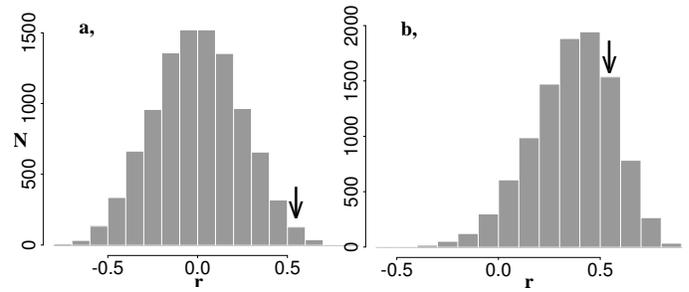}}
\caption{Histogram of correlation coefficients ($r$) of 10\,000 Monte-Carlo simulations
of 18  [Fe/H]$_{\rm spect}$ and
[Fe/H]$_{\rm phot}$ pairs.
a) No metallicity spread is assumed, the standard deviations of the data
are 0.22 and 0.13~dex, respectively. 
b) The data range over 0.34~dex in metallicity with 0.18 and 0.11~dex
standard
deviations.
Arrows indicate the regression coefficient between  the measured 
spectroscopic and photometric metallicities.
\label{fig3}
}
\end{figure}

The final sample of stars with [Fe/H] reliably determined  both
spectroscopically and photometrically  
consists of a limited number of 18 variables.
Fig.~\ref{fig2} shows these data, with the direct and inverse regression fits
overplotted.
The linear Pearson correlation coefficient between the two data sets is 0.54,   
 with 0.021 probability value. The small number of data
makes it difficult to estimate the statistical significance
of this result. Therefore, Monte-Carlo simulation has been performed
in order to decide whether
the correlation between the metallicities is real, or it is only
an artifact due to the large errors of the data.

The Monte-Carlo simulation was realized as follows.
Two artificial data sets were generated. First, 
the
0.22 and
0.13~dex
standard deviations of the [Fe/H]$_{\rm spect}$ and [Fe/H]$_{\rm phot}$  
data were supposed to arise  merely from the scatters of the observations
around a
single metallicity
value.
Secondly, it was assumed that a 0.34~dex [Fe/H] range was real, and the
standard deviations of the spectroscopic and photometric data were smaller,
0.18 and 0.11~dex, respectively (these values are the residual
scatters of the
regressions, see Fig.~\ref{fig2}). 
In both cases 10\,000 [Fe/H]$_{\rm spect}$ and [Fe/H]$_{\rm phot}$ data sets with 18
elements were generated.
The histograms of the correlation coefficients ($r$) of the simulated data
sets are shown
in 
Fig.~\ref{fig3}. While the observed $r=0.54$ correlation occurs marginally in
case a),
it has significant occurrence frequency in case b).
We can therefore draw the conclusion that in our sample the $r=0.54$
correlation is indeed significant.

As the two metallicity determinations 
are completely independent, the fact that a correlation between these quantities 
is found points to the reality of the observed metallicity spread.



\section{M$_{V}$ - [Fe/H] relation} 

\begin{figure}[t]
\resizebox{\hsize}{!}{\includegraphics{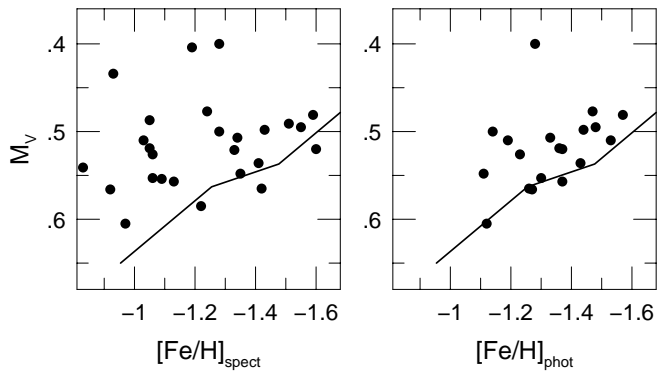}}
\caption{
Absolute magnitudes versus spectroscopic and photometric metallicities of 
M3 RRab stars.
$\mathrm{M_V = \langle V\rangle_{int} - 15.12}$~mag   \citep{h96}.
Lines indicate one of the theoretical ZAHB relations \citep{cal97}.
For the [Fe/H]$_{\rm phot}$ values at the different metallicities the
faintest stars (which are likely on the ZAHB) indicate a similar trend of
$\mathrm{M_V}$ with metallicity as expected, if the metallicity range is real.
The brightest star V60 has the longest period (0.77~d) and possibly is already in 
an evolved stage of the HB evolution. The second brightest among the spectroscopic 
sample is V202 which RR Lyrae classification is uncertain.
The less convincing evidence of an M$_{V}$ - [Fe/H] relation for the spectroscopic values
is possibly due to the larger errors in the [Fe/H]$_{\rm spect}$ values.
\label{fig4}}
\end{figure}

The metallicity dependence of the absolute magnitudes of horizontal
branch (HB) stars is well known, and is proven both from theoretical 
and observational points of view. Although the uniqueness, the slope
and the zero point of this relation are still controversial, 
the reality of the detected metallicity spread can be justified by 
checking the connection between the measured absolute magnitudes and 
metallicities of the variables. 

A strict connection between the luminosities and metallicities of HB stars
is only valid for zero age HB  (ZAHB) stars, as the luminosity of stars 
evolving off the ZAHB sequence increases, and this results a true internal 
scatter of any measured M$_{V}$ - [Fe/H] plot. The lower boundary of these 
plots have to follow the M$_{V}$ - [Fe/H] relation valid for ZAHB stars
within the limits of the observational uncertainties.

It is stressed that the brightness of the stars has no influence on 
either of the [Fe/H] determinations, which rules out  any a priori
connection between the values. If the metallicity spread of the variables is
real, than the faintest magnitudes at the different 
metallicities have to follow the global trend valid for ZAHB stars.

In Fig.~\ref{fig4} the absolute magnitudes
of the RRab sample are plotted versus their spectroscopic and photometric
metallicities. For the [Fe/H]$_{\rm phot}$ values, the faintest stars at  different
metallicities follow the trend predicted by ZAHB models closely.
Selecting ZAHB stars from the photometric sample by taking into account only the two 
faintest stars of each 0.1 dex [Fe/H] bins  a correlation with 
0.84 correlation coefficient and a 0.20 slope can be yield.
This slope  agrees extremely well with the model predictions which
typically yield slopes within the 0.18-0.26 range. 

The less convincing result for the spectroscopic metallicities is possibly due
to the larger errors in the spectroscopic data.

This result provides  further 
evidence that a  metallicity spread of about $0.3 - 0.4$~dex of the M3 RRab stars has 
been in fact detected.        

\section{Conclusion}

The statistically significant
correlation between the spectroscopic and photometric metallicities
of the RRab variables in M3, which have $\sim0.6$, and  0.4~dex ranges, respectively,
leads to the conclusion 
that a true spread in the metallicities of the variables exists. 
This is in contrast with the strictly homogeneous iron content of the cluster
red giants, though, it is also worth noting that in an early study CaI
overabundance of about 0.6 dex of a cluster giant (vZ1397) was found \citep{cohen}. 
This result has been neither confirmed nor rejected still today.
Although some systematic difference between the 
metallicity scales  of the red giants and the variables 
may exist,
the comparison of the observed metallicities 
indicates that the variables  show a metallicity range
spanning to about $0.3-0.4$ dex more metal rich values. 

Spectroscopic observations and chemical composition determinations
of cluster variables, apart from some $\Delta$S measurements showing
large scatter most probably due to observational errors,
exist  only of two clusters, $\omega$~Cen and M3.
In $\omega$~Cen as the iron abundances of the red giants have larger than 1.0 dex 
range, 
the result that the RR Lyrae metallicities are not homogeneous either
is not   surprising.
It is important to repeat, however, that no spectroscopic
observations of any other globular cluster variables exist.
The photometric metallicities as compiled in  \citet{kw} show unaccountable 
large scatter in many clusters where  no metallicity
spread of red giants is observed. This warns that a similar
result,  an observable metallicity spread of the variables 
might possibly occur in other clusters as well.

\begin{table}[!ht]
\caption[]{Mean meatallicity values of M3 variables within smaller and larger radial
distances }\label{rad}
\begin{tabular}{ccccccc}
\hline
\noalign{\smallskip}
 &$\langle{\rm [Fe/H]}_{\rm spect}\rangle$ & $\sigma$ & N &
$\langle{\rm [Fe/H]}_{\rm
phot}\rangle$& $\sigma$ & N \\
\noalign{\smallskip}
\hline
 \noalign{\smallskip}
 $R<2\farcm 5$ &  $ -1.08$ &0.23& 3  &$-1.11$&0.01& 2\\
 $R<3\farcm 5$ &  $ -1.13$ &0.18& 8  &$-1.24$&0.14& 5\\
 $R>3\farcm 5$ &  $ -1.28$ &0.22& 20 &$-1.36$&0.13& 14  \\
 all&             $ -1.24$ &0.22& 28 &$-1.33$&0.14& 19 \\
\noalign{\smallskip}
\hline
\noalign{\smallskip}
\end{tabular}
\end{table}

\citet{cc01} have found that there are 12 RR Lyrae stars in the inner 81" region lying 
$0.1-0.2$~mag below the ZAHB. They excluded the possibility that
these stars were too faint because of photometric
inaccuracies, however, they could not give any 
plausible reason for their low luminosities.
As shown in Fig.~\ref{fig4} metallicity effect may partly explain the
faintness of these stars.
The variables discussed in the present paper
although do not show clear evidence of correlation between their metallicities and  
radial distances from the cluster center, the innermost stars have
systematically larger metallicities than the outer variables  as is shown in 
Table~\ref{rad}.

Radial differences of the HB type and of the frequency of blue straggler stars (BSS)
have been  also detected in M3, namely the  horizontal branch of the inner region is 
bluer and its BSS population is more numerous \citep{catelan,ferr}.
Though these phenomena still lack explanation,
all of them indicate that there is a dynamical segregation 
within the cluster. Fitting the luminosity function of M3, \citet{rood}
have also found that the lower mass stars ($M<0.8M_{\sun}$; stars below the turnoff point)
 do not have a uniform radial distribution as well.
If the stellar composition, and central density of the
stars nearer to the cluster's center are in some degree different,
this might explain our finding, supposing that all the spectroscopically
measured cluster red giants with homogeneous metallicities belong to 
the outskirt of the cluster.

If the iron content is indeed not homogeneous in HB stars,
its explanation could probably be found in the early
star formation and chemical evolution history of the system.
Recent investigations indicate that self-pollution
could have been efficient  
during the first some hundred million years of the cluster lifetime, 
just before its first crossing of the galactic plane, 
which removed all the intracluster gas from the cluster.
As a manifestation of self-pollution, both the then  existing stars   accreted 
processed material, and  the composition of the intracluster gas
which  formed  to stars later, was altered \citep{thoul,danton}.
Models show that in M3,  $11\%$ of the mass of a 1~$M_{\sun}$ star
may originate from the mass loss of the first generation, large mass, evolved stars, 
and not from the uncontaminated protocluster gas according to the accretion scenario
\citep{thoul}. 
\citet{shust} modeled supernova explosions in globular clusters,
and concluded that  self-pollution 
via SN ejetion could also happen in globular 
clusters, that might be a clue to explain any detected metallicity
spread. They also argue that SN self-pollution results in smaller scatter of the 
observed iron abundances than in any of the lighter elements.
This helps to interpret why [Fe/H] values seem to be homogeneous in most of the
globular clusters.

The question why [Fe/H] dispersion only of RR Lyrae stars is detected,
while red giants seem to be uniform in their iron content
is difficult to answer  on the bases of the presently available observational data.
Because of the small sample 
size and the large uncertainties especially in the spectroscopic data
of RR Lyrae stars,
good quality detailed spectroscopic measurements of a much larger RR Lyrae 
 sample is needed to confirm and to find the correct interpretation of the
suggested metallicity dispersion of RR Lyrae stars in M3.
Similar studies of other globular clusters are also  encouraged.

\begin{acknowledgements}
This research has made use of the SIMBAD database, operated at CDS 
Strasbourg, France. This work has been supported by OTKA grant T30954.
\end{acknowledgements}

\bibliographystyle{aa}

\begin{thebibliography}{}
\bibitem[Bakos \& Jurcsik(2000)]{bj00} Bakos, G.,  and Jurcsik, J. 2000, ASPCS Vol. 203, 255
\bibitem[Benk\H o et~al.(2003)]{bj} Benk\H o, J. et al. 2003, in preparation
\bibitem[Bell \& Dickens(1980)]{bell} Bell, R. A., and Dickens, R. J. 1980, \apj, 242, 657
\bibitem[Bernard(1982)]{bern} Bernard, A. 1982, \pasp, 94, 700
\bibitem[Caloi et al.(1997)]{cal97} Caloi, V, D'Antona, F., and Mazzitelli, I. 1997, \aap, 
320, 823
\bibitem[Carretta et~al.(1998)]{c98} Carretta, E., Cacciari, C., Ferraro, F. R., Fusi Pecci, F., and Tessicini, G. 1998, \mnras, 298, 1005
\bibitem[Catelan et~al.(2001)]{catelan} Catelan, M., Ferraro, E. R., and Rood, R. T. 2001, \apj, 560, 970
\bibitem[Cavallo \& Nagar(2000)]{cav} Cavallo, R. M., and Nagar, N. M. 2000, \aj, 120, 1364
\bibitem[Cohen(1978)]{cohen} Cohen, J. 1978, \apj, 223, 487
\bibitem[Corwin \& Carney(2001)]{cc01} Corwin, T. M., and Carney, B. W. 2001, \aj, 122, 3183
\bibitem[D'Antona et al.(2002)]{danton} D'Antona, F., Caloi, V., Montalb\'an, J., Ventura, P., and Gratton, R.
2002, \aap, 395, 69 
\bibitem[Ferraro et al.(1997)]{ferr} Ferraro, F. R., Paltrinieri, B., Fusi Pecci, F.,
Cacciari, C., Dorman, B., Rood, R. T., Buonanno, R., Corsi, C.E., Burgarella, D., and Laget,
M. 1997, \aap, 324, 915 
\bibitem[Harris(1996)]{h96} Harris, W. E. 1996, \aj, 112, 1487,
http://physun.physics.mcmaster.ca/Globular.html
\bibitem[Jurcsik(1998)]{j98} Jurcsik, J. 1998, \aap, 333, 571
\bibitem[Jurcsik \& Kov\'acs(1996)]{jk96} Jurcsik, J., and  Kov\'acs, G. 1996, \aap, 312, 111
\bibitem[Kaluzny(1998)]{k98} Kaluzny, J., Hilditch, R. W., Clement, C., and Rucinski, S. M. 1998, \mnras, 296, 347
\bibitem[King et~al.(1998)]{king98} King, J. R., Stephens, A., Boesgaard, A. M., and
Deliyannis, C. 1998, \aj, 115, 
666
\bibitem[Kov\'acs \& Walker(2001)]{kw} Kov\'acs, G., and Walker, A. 2001, \aap, 374, 264
\bibitem[Kraft et~al.(1992)]{kraf} Kraft, R. P., Sneden, C., Langer, G. E., and Prosser, C. 1992, \aj, 104, 645 
\bibitem[Kraft \& Ivans(2002)]{kraft} Kraft, R. P., and Ivans, I. I. 2003, \pasp, 115,...
\bibitem[Lambert et~al.(1996)]{la96} Lambert, D. L., Heath, J. E., Lemke, M., and Drake, J. 1996, \apjs, 103, 183
\bibitem[Langer et~al.(1998)]{lang98} Langer, G. E., Fischer, D., Sneden, C., and Bolte, M. 1998, \aj, 115, 685
\bibitem[Layden(1994)]{layden} Layden, A. 1994, \aj, 108, 1016
\bibitem[Rood et al.(1999)]{rood} Rood, R. T., Carretta, E., Paltrinieri, B., Ferraro,
F.R., Fusi Pecci, F., Dorman, B., Chieffi, A., Straniero, O., and Buonanno, R. 1999, \apj,
523, 752
\bibitem[Sandstrom et~al.(2001)]{s01} Sandstrom, K., Pilachowski, C. A., and Saha, A. 2001, \aj, 122, 3212
\bibitem[Shustov \& Wiebe(2000)]{shust} Shustov, B. M., and Wiebe, D. S. 2000, \mnras, 319,
1047
\bibitem[Th\'evenin \& Idiart(1999)]{ti99} Th\'evenin, F., and Iidart, T. P. 1999, \apj, 521, 753
\bibitem[Thoul et al.(2002)]{thoul} Thoul, A. Jorissen, A., Goriely, S., Jehin, E., Magain, P., Noels, A., and
Parmentier, G. 2002, \aap, 383, 491
\end{thebibliography}

\end{document}